# The Art of Data Hiding with Reed-Solomon Error Correcting Codes

### Fredrick R. Ishengoma
Computer Engineering and
Applications
The University of Dodoma
P.O. Box 490, Dodoma,
Tanzania

## ABSTRACT
With the tremendous advancements in technology and the Internet, data security has become a major issue around the globe. To guarantee that data is protected and does not go to an unintended endpoint, the art of data hiding (steganography) emerged. Steganography is the art of hiding information such that it is not detectable to the naked eye. Various techniques have been proposed for hiding a secret message in a carrier document. In this paper, we present a novel design that applies Reed-Solomon (RS) error correcting codes in steganographic applications. The model works by substituting the redundant RS codes with the steganographic message. The experimental results show that the proposed design is satisfactory with the percentage of decoded information 100% and percentage of decoded secret message 97.36%. The proposed model proved that it could be applied in various steganographic applications.

## Keywords
Reed-Solomon Error Correcting Codes

## 1. INTRODUCTION
Security of the information is one of the crucial factors in data transmission [1]. One of the approaches used to ensure reliable transmission of confidential information is steganography. Steganography is the art and science of invisible communication [2]. This is accomplished through hiding secret message in other information, thus hiding the existence of the communicated information [3]. Steganography focuses on keeping the existence of a message secret [4]. Usually it is done in such a way that it is hard for an opponent to notice the presence of a secreted message in the transmitted data. Data transmitted appear visible, risk-free and normal.

Two of the significant aims of steganography in digital media are (i) to guarantee data integrity and (ii) to deliver proof of the copyright. Therefore, data should stay hidden in a host signal, even if that signal is subjected to manipulation such as filtering, resampling, cropping, or lossy data compression [5]. One way of implementing steganography is by developing a solution that makes use of a communication channel.

Steganographic schemes based on Error Correcting Codes (ECCs), (The Bose, Chaudhuri, and Hocquenghem (BCH) codes, hamming distance) have been previously studied and explored [6,7]. However, the existing approaches of ECCs in steganography are faced with various shortcomings that include the size of the secret message, decoding algorithm being too complex, and the extent of distortion [8,9,10]. This presents the needs for further research in the application of error correcting codes in steganography. Innovating new techniques in steganography will assist in overcoming the existing shortcoming and present new opportunities of steganography application [8]. This research study employs Reed-Solomon ECCs in steganographic application and explores its performance.

## 2. STEGANOGRAPHY
Steganography requires the following attributes:

1) Carrier document($C$): Also known as cover media. It holds the hidden data. This includes text, pictures, audio files, videos, IP datagram, documents and other file types.
2) Secret Message($M$): This is the hidden data in form of plain or cypher text.
3) Stego function ($Fe$) and stego inverse function ($Fe^{-1}$).
4) Key ($K$) for hiding and unhide the message.

Figure 1 depict the typical steganographic operation whereby $Fe$ works on and ($C$) and ($M$) along with ($K$) to create a stego media ($S$).

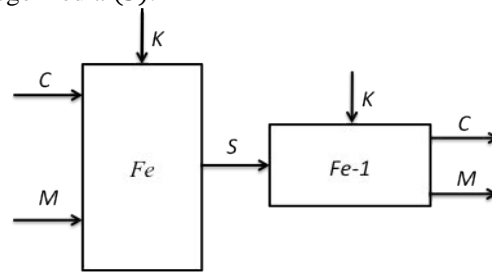

**Fig 1: Typical steganography scenario**

There are several steganography approaches that include text steganography, audio steganography, video steganography, image steganography and protocol steganography. Taking example of audio steganography. In audio steganography, secret message (M) is embedded into the digital audio signal. This cause a slightly variation of the binary audio sequence. Digital audio is discrete compared to continuous signal in analog audio. A discrete audio signal is generated by sampling a continuous analog signal at a definite rate. For instance, in CD audio the standard sampling rate is around 44 kHz.

The digital audio is saved in a computer in a form of 0s and 1s. By using least significant bit (LSB) encoding, binary samples of the digital audio are replaced by the binary equivalent of the secret message (M). Suppose we are to embed letter 'A' (01100101 in binary) to an mp3 file with each sample 16 bits then LSB of 8 successive samples (every one with 16 bits) is substituted with every bit of binary correspondent of letter 'A'.

**Table 1. Example of embedding letter 'A' to an mp3 file**

| MP3 Audio Sample | 'A' | MP3 Audio Sample with encoded data |
|---|---|---|
| 1001 1000 0011 1100 | 0 | 1001 1000 0011 110**0** |
| 1101 1011 0011 1000 | 1 | 1101 1011 0011 100**1** |
| 1011 1100 0011 1101 | 1 | 1011 1100 0011 110**1** |
| 1011 1111 0011 1100 | 0 | 1011 1111 0011 110**0** |
| 1011 1010 0111 1111 | 0 | 1011 1010 0111 111**0** |
| 1111 1000 0011 1100 | 1 | 1111 1000 0011 110**1** |
| 1101 1100 0111 1000 | 0 | 1101 1100 0111 100**0** |
| 1000 1000 0001 1111 | 1 | 1000 1000 0001 111**1** |

## 3. CODING THEORY

Coding theory is the study of the transmission of data across a noisy channel, with the ultimate goal of successfully recovering from an error-ridden received message back to its original in error-free form [11]. Data is transmitted in form of bits (0s and 1s) over channels (air, space, wire), whereby the presence of noisy is common. The noise may alter the message sent and the receiver may not get the same message sent from the sender. Error-correcting codes were introduced with the purpose of ensuring error-free data transmission over noisy communication channels. The major challenge in coding theory is to transmit data over a noisy channel with a highest possible reliability.

### 3.1 Hamming Distance

Hamming distance [12] is used to measure how much two bit strings differ. The hamming distance $d\,(x, y)$ between two bit strings $x = x_1 x_2 \dots x_n$ and $y = y_1 y_2 \dots y_n$ is the number of positions in which these two strings differ.

$$d_H\,(x, y) = \sum_{j=0}^{n-1} W_H\,(x_j + y_j),$$

$$where\ (x_j + y_j) = \begin{cases} 0, & x_j = y_j \\ 1, & x_j \neq y_j \end{cases} \tag{1}$$

$$d_H\,(x, y) = W_H\,(x, y) \tag{2}$$

### 3.2 Reed-Solomon Codes

Reed-Solomon (RS) codes are one of Forward-Error Correction (FEC) codes that was developed in 1960 by Irving S. Reed and Gustave Solomon in 1960 [11]. RS codes works in a way that, if in communication a block of data is lost or totally erased, as long as sufficient of the residual blocks are received, it is likely to recover the lost data. In RS codes, a block of $k$ bits of data is encoded into $n$ bits of data (codeword) where $n > k$ for data transmission [13].

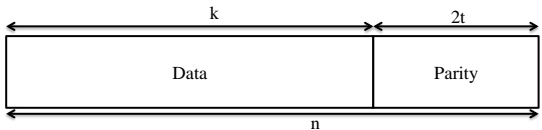

**Fig 2: RS Codeword**

Figure 2 depict RS codeword specified as RS $(n, k)$ where $n$ is the size of the codeword, $k$ is the number of data symbol and $2t$ is the number of parity symbols with each symbol containing s number of bits.

### 3.2.1 Encoding

The study considers RS code being a specific case of BCH codes of length $n = q - 1$. Consider a data polynomial $d(x) = d_0 + d_1 x + \dots + d_{k-1} x^{k-1}$ under Galois Field (GF) with each $d_i \epsilon\, GF(q)$. *Cauchy matrix* [14] was used for encoding. A codeword $c$ is encoded by adding $n - k$ parity check symbols to the data.

$$c = (c_0, c_1, \dots c_{n-1}) \tag{3}$$

$$= (p_0, p_1, \dots, p_{n-k-1}, d_0, d_1, \dots d_{k-1}) \tag{4}$$

Generator Matrix: $G = (I \quad A)$, where $I = k\ x\ k$ identity matrix and $A = k\ x\ (n - k)$ Cauchy matrix.

$$A_{i,j} = \frac{u_i, v_j}{x_i + y_j}, 0 \leq i \leq k - 1, 0 \leq j \leq n - k - 1 \tag{5}$$

$$x_j = \alpha^{n-1-i}, 0 \leq i \leq k - 1 \tag{6}$$

$$y_j = \alpha^{n-1-k-j}, 0 \leq j \leq n - k - 1 \tag{7}$$

$$u_i = \frac{1}{\prod_{0 \leq l \leq k-1}(\alpha^{n-1-i} - \alpha^{n-1-l})}, 0 \leq i \leq k - 1 \tag{8}$$

$$v_j = \prod_{0 \leq l \leq k-1}(\alpha^{n-1-k-j} - \alpha^{n-1-l}), 0 \leq j \leq n - k - 1 \tag{9}$$

Where Parity check vector $p = (p_0 \dots p_{n-k-1})$:

$$p = dA \tag{10}$$

$$p = \sum_{i=0}^{k-1} d_i\, A_{i,j} \tag{11}$$

$$p = v_j \sum_{i=0}^{k-1} \frac{d_i, u_i}{x_i + y_j}, 0 \leq j \leq n - k - 1 \tag{12}$$

### 3.2.2 Decoding

*Syndrome-Based Decoding* [15, 16] was used in this study. Error in a communication channel is defined as: $e(x) = e_0 + e_1 x + \dots + e_{n-1} x^{n-1}$ (13)

Consider decoder input to be:

$$v(x) = c(x) + e(x) = v_0 + v_1 x \dots v_{n-1} x^{n-1} \tag{14}$$

Where $v_i$ = codeword's $i$ [th] element. From roots of the Generator Polynomial (GP), the polynomial is examined.

$$v(\alpha^j) = c(\alpha^j) + e(\alpha^j) \tag{15}$$

$$v(\alpha^j) = e(\alpha^j) = \sum_{i=0}^{n-1} e_i \alpha^{ij}, j = 1, 2, \dots 2t \tag{16}$$

Using Fourier transformation, $j$ [th] syndrome ($s_j$) is:

$$v(\alpha^j) = \sum_{i=0}^{n-1} v_i \alpha^{ij}, j = 1, 2, \dots 2t \tag{17}$$

If $0 \leq v \leq t$ errors occur in unknown location $i_1, i_2, \dots i_v$ of a communication channel, then:

$$e(x) = e_{i_1} x^{i_1} + \dots + e_{i_v} x^{i_v} \tag{18}$$

Finding the $i$'s and the $e$'s:

Let $Y_l = e_{i_l}$ for $l = 1,2,\dots v$ and $\alpha^{i_l}$ for $l = 1,2,\dots,v$. After syndromes evaluation, equations with $v$ unknowns are obtained:

$$S_1 = Y_1 X_1 + Y_2 X_2 + \cdots Y_v X_v \qquad (19)$$

$$S_j = Y_1 X_1^j + Y_2 X_2^j + \dots + Y_v X_v^j \qquad (20)$$

$$S_{2t} = Y_1 X_1^{2t} + Y_2 X_2^{2t} + \dots + Y_v X_v^{2t} \qquad (21)$$

Finding the solution with the least amount of error completes the RS decoding process

## 4. PROPOSED MODEL

The proposed model of applying steganography in Reed Solomon communication channel works by substituting some of the redundant ECCs with the secret message. The assumption made under the proposed model is that RS has enough capability such that even if some of its ECCs are substituted with the secret message it will still have no effect on its performance. Some locations of the redundant RS codes are selected and the secret message is substituted. In this model, the secret message is embedded in RS code as errors exclusive of affecting the RS encoder or decoder.

Reed-Solomon encoder creates the codeword by taking input steganographic data. The positions in the codeword $C(x)$ that are to be used for data hiding are indicated by using the powers of α. Then the data in the indicated positions are substituted with the secret message resulting into a modified codeword/steganographic codeword $C(x, S_M)$. The modified codeword is conveyed in a communication channel that is disposed to errors and reached as a received codeword RC (x). The received codeword is then decoded by RS codes to extract the error correcting bits and also by steganographic decoder to extract secret message.

| Algorithm 1: Proposed Steganography with RS Codes |
|---|
| *Input: Codeword, Secret Message, Data* |
| *Output: Secret Message* |
| *//$D_i$: Normal data to be transmitted by RS* |
| *//RS: Reed-Solomon Codes* |
| *//C(x): Codeword* |
| *//P (x): Positions in C (x) for data hiding* |
| *//DP (x): Data in P (x)* |
| *//$S_M$: Steganographic Message* |
| *//SD: Steganographic decoder* |
| *//C (x, $S_M$): Modified Codeword with Secret Message* |
| *// $D_A$: Initial location of the C (x, $S_M$)* |
| *// $D_A$: C (x, $S_M$) Destination Location* |
| *1. RS generate C (x)* |
| *2. Indicate P (x) in C (x) for data hiding* |
| *3. Substitute DP (x) with $S_M$* |
| *4. Generate C (x, $S_M$)* |
| *5. Transmit C(x, $S_M$) from $D_A$ to $D_B$* |
| *6. Decode C(x, $S_M$) with RS* |
| *7. Extract $D_t$* |
| *8. Decode C(x, $S_M$) with SD* |
| *9. Extract Secret Message* |

## 5. PERFORMANC EVALUATION

This study applied experimental design. Experimental design deals with analysis of data produced from a series of experiments. An application using Java-OOP for simulating the proposed steganography application under RS code was programmed using Java-OOP. The experiment used RS (31,19) with the ability of correcting maximum of 4 symbols.

2 bits of $S_M$ were embedded in the codeword.

Metrics used in the experiment are (i) percentage of decoded information (%$D_i$) and (ii) percentage of decoded secret message (%$DS_M$). %$D_i$ signifies if the embedded secret message has effect to the RS decoding capability. %$DS_M$ signifies how much has the steganographic message been successful decoded. %$D_i$ and %$DS_M$ were determined by performing 100 tests. The experiment was tested in accordance to single error.

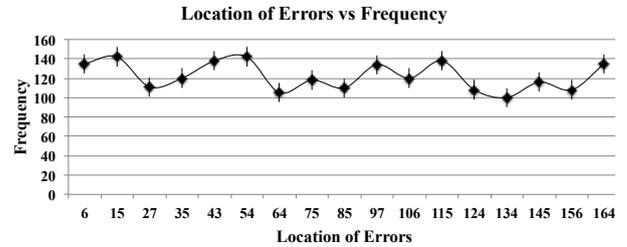

**Fig 3: The graph of location of errors vs. frequency**

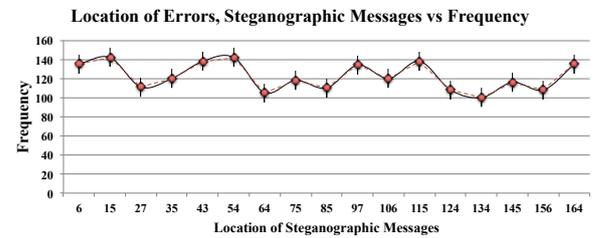

**Fig 4: The graph of location of steganographic messages vs. frequency**

In the experiment, random steganographic message was encoded into data and codeword was generated by RS-ECC. Random bits were altered to replicate the noisy communication channel. Information was received and decoded by RS-ECC decoder and compared with the sent information.

The experimental results found that %$D_i$ was 100% and %$DS_M$ was 97.96. Steganographic methods needs that the distribution of steganographic data to be identical in order to be resistible to steganographic analysis. Therefore, the experimental results prove that the proposed technique is satisfactory enough to be applied in steganographic applications.

Furthermore, the proposed model was validated under burst error environment. In the experiment, in every codeword, arbitrarily locations were chosen that are with 6 bits burst error length. The experimental results were very similar to single bit experimental results with %$D_i$ is 100% and %$DS_M$ is 96.5.

**Table 2: Experimental results under burst mode**

| RS Code | No. Tests | %$D_i$ | %$DS_M$ |
|---|---|---|---|
| RS (31,19) | 100 | 100 | 96.5 |

For the whole two experiments (single bit and burst errors) the capacity of the RS codes were not surpassed, thus minimizing the probability of suspicion of steganographic presence.

## 6. CONCLUSIONS

In this paper a new model of applying Reed-Solomon ECC into steganographic application was proposed. The model works by substituting redundant codes with the secret message. The experimental results proved the model is working successfully. The proposed model could be applied in a wide range of application including covert communications, coded anti-piracy, intelligence services and even illegally example by terrorist communication. Future work will focus on the extension of the proposed model by considering the size of secret message and extent of distortion.